\documentclass[aps,pre,reprint,groupedaddress]{revtex4-1}

\usepackage{dcolumn}
\usepackage{bm}
\usepackage{amsmath}
\usepackage{color}
\usepackage{graphicx}

\begin{document}

\title{
Universal L\'evy's stable law of stock market and its characterization
}

\author{Takumi Fukunaga}
\email{fukunaga.takumi.66v@st.kyoto-u.ac.jp}
\author{Ken Umeno}
\email{umeno.ken.8z@kyoto-u.ac.jp}

\affiliation{
Department of Applied Mathematics and Physics,\\
Graduate School of Informatics, Kyoto University.
}

\date{\today}

\begin{abstract}
L\'evy's stable distribution is a probability distribution with power-law tails and skewness, which are the essentially important characters in financial analysis. Although the Gaussian distribution has long been used in financial modelings, we suggest that L\'evy's stable distribution is more suitable by theoretical reasons and analysis results. When L\'evy's stable distribution is fitted to the stock market, its parameters showed characteristic values. In long-term analysis, similar values were obtained for four different stock indices, which seem to be universal. In short-time analysis, however, when looked at on a daily basis, the parameters fluctuate in correlation with stock prices. We conclude that L\'evy's stable distribution could be a following probability distribution in the area of finance.
\end{abstract}


\maketitle

\section{Introduction}

The elucidation of mechanisms of stock markets is a long-standing problem. Since the financial crisis in 2008, which had occurred from the bankruptcy of Lehman Brothers, the revision of classical statistic approaches to deal with financial markets are receiving attentions.

In the analysis, the shape of the one-time probability distribution of prices is important in the quantitative description of financial markets, where the Gaussian distribution is widely used for such as the Black-Scholes model \cite{bs1973} for pricing derivatives. This Gaussianity assumption of the price fluctuation has been used for many years not only because of its simpleness in analysis but also because of the prevailing view that the central limit theorem could be applied to the price fluctuation \cite{markowitz1952,merton1971}. Recently, however, this assumption has been questioned. As an example, many price fluctuations are observed which cannot be represented with the Gaussian distribution and having power-law tails, first found by Mandelbrot on cotton prices \cite{mandelbrot1963}. Therefore, new methods for analyzing financial markets is needed using not the Gaussian distribution but an alternative.

One of the most famous power-law tailed distributions is L\'evy's stable distribution. There have been several works which applied L\'evy's stable distribution to stock markets \cite{fama1965,miller1974,mantegna1995,liu1999,gopikrishnan1999,plerou1999,plerou2000,gopikrishnan2000,rachev2000,stanley2003}. In these investigations, they pointed out that L\'evy's stable distribution fits better than the Gaussian distribution to financial markets. It is still debatable whether L\'evy's stable distribution is applicable, since there is not enough theoretical background and there is not a universal analyzing method for estimating parameters of L\'evy's stable distribution.

It is not possible to apply ordinary analyzing methods to L\'evy's stable distribution  due to its peculiar property. Non-parametric approaches to estimate L\'evy's stable parameters, such as the log-log linear regression and the Hill estimator, have limited range of estimation since the error increases near $\alpha=2$ \cite{weron2001}. Other representative methods for estimating L\'evy's stable parameters are the quantiles method \cite{fama1971,mcculloch1986}, the fractional lower order moment method \cite{nikias1993,nikias1995}, the logarithmic moment method \cite{kuruoglu2001} and the characteristic function method \cite{koutrouvelis1980}. While the former three methods have some constraints of restricted ranges of parameters, high computational cost, or requiring large number of data, the characteristic function method seems to be the most appliable. We would like to use a simple approach based on the characteristic function, which is tested to be valid and clears the above issues.

In this paper, we analyze log-returns of stock markets by modeling with L\'evy's stable distribution, and conclude that L\'evy's stable distribution is applicable to stock markets by the theoretical background and the analysis result.

\section{Properties of L\'evy's stable distribution}

L\'evy's stable distribution is expressed by its characteristic function $\phi(k)$, while a general expression for the probability density function $f(x)$ does not exist. These are related as
\begin{eqnarray}
f(x)=\frac{1}{2{\rm \pi}}\int^{\infty}_{-\infty}\phi(k){\rm e}^{-{\rm i}xk}{\rm d}k,
\label{eq:1}
\end{eqnarray}
where
\begin{eqnarray}
\phi(k)={\rm exp}\left\{{\rm i}\delta k-|\gamma k|^{\alpha}\left[1+{\rm i}\beta{\rm sgn}(k)\omega(k,\alpha)\right]\right\}
\label{eq:cf_phi}
\end{eqnarray}
and
\begin{eqnarray}
\omega(k,\alpha)&=&
\left\{
\begin{array}{ll}
{\rm tan}\frac{\pi\alpha}{2} & (\alpha\neq1)\\
\frac{2}{\pi}{\rm log}|k| & (\alpha=1).
\end{array}
\right.
\label{eq:omega}
\end{eqnarray}

Here, $\alpha\in(0,2]$ is the stability parameter, $\beta\in[-1,1]$ the skewness parameter, $\gamma\in[0,\infty)$ the scale parameter and $\delta\in(-\infty,\infty)$ the location parameter. The case of $(\alpha,\beta)=(2,0)$ corresponds to the Gaussian distribution and $(\alpha,\beta)=(1,0)$ to the Cauchy distribution.

L\'evy's stable distribution has power-law tails. When $|x|\rightarrow\infty$ where $c_+\in(0,\infty)$ and $c_-\in(0,\infty)$, the following equation holds:
\begin{eqnarray}
f(x)=
\left\{
\begin{array}{ll}
c_+x^{-(1+\alpha)} & (x\rightarrow\infty)\\
c_-|x|^{-(1+\alpha)} & (x\rightarrow-\infty).
\end{array}
\right.
\label{eq:power}
\end{eqnarray}

In addition to the fact that the probability density function cannot be represented explicitly, the mean cannot be defined for $\alpha\in(0,1]$ and the variance diverges for $\alpha\in(0,2)$.

\section{Generalized central limit theorem}

The classical central limit theorem states that the sum of independent and identically distributed random variables with a finite mean and variance converges to the Gaussian distribution. According to the generalized central limit theorem \cite{gclt1954}, when the variables follow the power-law of Eq. (\ref{eq:power}) with an infinite variance, the sum of variables converges to L\'evy's stable distribution with the stability parameter $\alpha$ and the skewness parameter $\beta=(c^+-c^-)/(c^++c^-)$. Further, a theorem proposed as the super generalized central limit theorem \cite{shintani2017} states that the sum of independent (not necessarily identically) distributed random variables with an infinite variance also converges to L\'evy's stable distribution.

Here, if $t$ is time and $S(t)$ a stock price, then its log-return $X(t):={\rm log}\left\{S(t)/S(0)\right\}$ is known to follow a probability distribution with power-law tails. Besides, the stock indices are composed of the sums of various stock prices. Thus, it can be suggested that L\'evy's stable distribution is more suitable than the Gaussian distribution for analyzing the stock indices. 

\section{Estimation of L\'evy's stable parameters}

When we analyze data, we often assume that they are {\it ergodic}. In general, if random variables $X_n$ $(n=1,2,\dots)$ are ergodic with the integrable function $f(x)$, the preserving map $T(x)$ and the measure $\rho(x){\rm dx}$ in the space $M$, then the following equation holds \cite{halmos1956,nemytskii1960}:
\begin{eqnarray}
\lim_{N\to\infty}\frac{1}{N}\sum^N_{n=1}f(T^nx)=\int_Mf(x)\rho(x){\rm d}x.
\label{eq:ergodicity}
\end{eqnarray}
Then, to consider characteristic functions, Eq. (\ref{eq:ergodicity}) comes out to be the following ergodic equality \cite{umeno2016}:
\begin{eqnarray}
\lim_{N\to\infty}\frac{1}{N}\sum^N_{n=1}{\rm exp}({\rm i}kX_n)=\int^{\infty}_{-\infty}{\rm exp}({\rm i}kx)f(x){\rm d}x,
\end{eqnarray}
from which we have
\begin{eqnarray}
\phi(k)=\lim_{N\to\infty}\frac{1}{N}\sum^N_{n=1}{\rm exp}({\rm i}kX_n)
\end{eqnarray}
which is consistent with Eq. (\ref{eq:1}). This assumption is usually used in data analysis especially in statistics, and it allows us to empirically obtain the probability distribution. Hence, the empirical characteristic function $\phi_N(k)$ of a large number of data set $X_n$ $(n=1,2,\dots,N)$ can be calculated as
\begin{eqnarray}
\phi_N(k)=\frac{1}{N}\sum^N_{n=1}{\rm exp}({\rm i}kX_n).
\end{eqnarray}
When the data follow L\'evy's stable distribution with the parameters $(\alpha,\beta,\gamma,\delta)$ ($\alpha\neq1$, $k>0$), the characteristic function $\phi_N(k)$ can be represented as
\begin{eqnarray}
\phi_N(k)={\rm exp}\left[{\rm i}\delta k-(\gamma k)^{\alpha}\left(1+{\rm i}\beta{\rm tan}\frac{{\rm \pi}\alpha}{2}\right)\right]
\label{eq:phi_estimate}
\end{eqnarray}
from Eqs. (\ref{eq:cf_phi}) and (\ref{eq:omega}).

In order to estimate L\'evy's Stable parameters, we use the approach based on Koutrouvelis's method \cite{koutrouvelis1980} with an added process of {\it normalization} which is discussed later. With Eq. (\ref{eq:phi_estimate}), we can derive
\begin{eqnarray}
{\rm log}\left(-{\rm log}\left|\phi_N(k)\right|\right)=\alpha{\rm log}k+\alpha{\rm log}\gamma
\label{eq:alpha_gamma}
\end{eqnarray}
and
\begin{eqnarray}
\frac{1}{k}{\rm arctan}\frac{\phi_{N,{\rm I}}(k)}{\phi_{N,{\rm R}}(k)}
=
-\beta\gamma^{\alpha}{\rm tan}\frac{\pi\alpha}{2}k^{\alpha-1}+\delta,
\label{eq:beta_delta}
\end{eqnarray}
where each of $\phi_{N,{\rm I}}(k)$ and $\phi_{N,{\rm R}}(k)$ corresponds to the real and imaginary part of $\phi_N(k)$. The parameters $(\alpha,\beta,\gamma,\delta)$ can be estimated by the linear regression method in Eqs. (\ref{eq:alpha_gamma}) and (\ref{eq:beta_delta}) around $k=0$. If $(\gamma,\delta)$ are far from the standard value of $(1,0)$, then it is not possible to estimate each parameter accurately. In this case, the data should be normalized to $(\gamma,\delta)=(1,0)$ and then estimate $(\alpha,\beta)$.

While the standard estimation method use the probability density function from the actual data with difficulty in estimating the tails of the distribution which are essentially important part of L\'evy's stable distribution, the present method here can detect the tail through the characteristic function. In addition, this method has a faster convergence according to the increasing number of data. Thus, we can analyze with relatively few data.

\section{Application to stock markets}

In this paper, we analyze four stock indices: Nikkei 225 (from 1/4/1978 to 5/31/2017), S\&P 500 (from 1/2/1975 to 5/31/2017), Dow 30 (from 1/2/1986 to 5/31/2017) and SSEC (from 1/5/1998 to 5/31/2017). These data were downloaded from Yahoo Finance (\url{http://finance.yahoo.com}). For each stock index, the daily log-returns are calculated at first, and then the parameters $(\alpha,\beta)$ are estimated after the standardization.

\begin{table}[tb]
\caption{\label{tab:stock}Estimated parameters $(\alpha,\beta)$ from four stock indices.}
\begin{ruledtabular}
\begin{tabular}{cccc}
\textrm{Stock index}&
\textrm{Number of data}&
\textrm{$\alpha$}&
\textrm{$\beta$}\\
\colrule
Nikkei 225 & 9934 & $1.570$ & $-0.162$\\
S\&P 500 & 10698 & $1.638$ & $-0.127$\\
Dow 30 & 7918 & $1.604$ & $-0.146$\\
SSEC & 4691 & $1.573$ & $-0.072$\\
\end{tabular}
\end{ruledtabular}
\end{table}

Table \ref{tab:stock} shows that the daily log-returns of {\it all} four considered stock markets are well fitted by a L\'evy's stable distribution with $\alpha\approx1.6$, which is remarkably different from the Gaussian distribution where $\alpha=2$, and consistent with Mandelbrot's findings on cotton prices of more than 50 years ago, where $\alpha=1.63$ \cite{mandelbrot1963}.

\begin{figure}[tb]
\begin{tabular}{cc}
\begin{minipage}{0.5\hsize}
\includegraphics[width=4.3cm,bb=0 0 767 598]{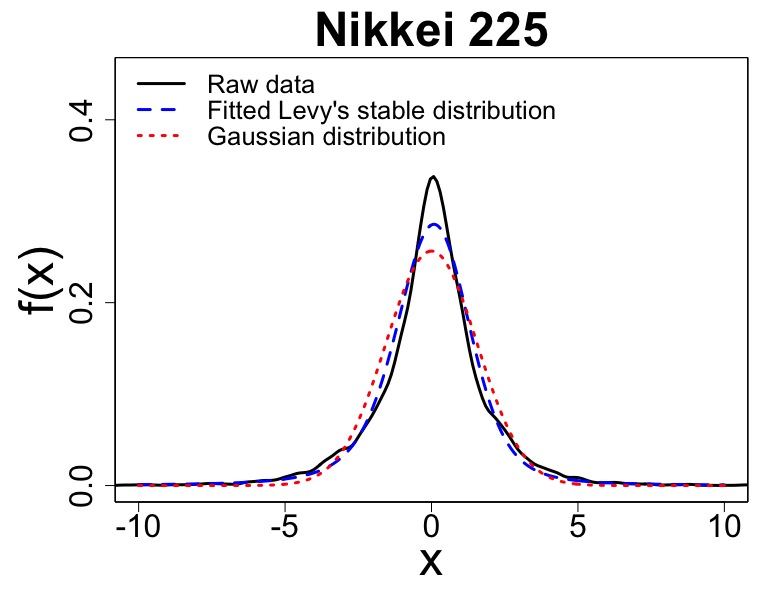}
\end{minipage}
\begin{minipage}{0.5\hsize}
\includegraphics[width=4.3cm,bb=0 0 767 598]{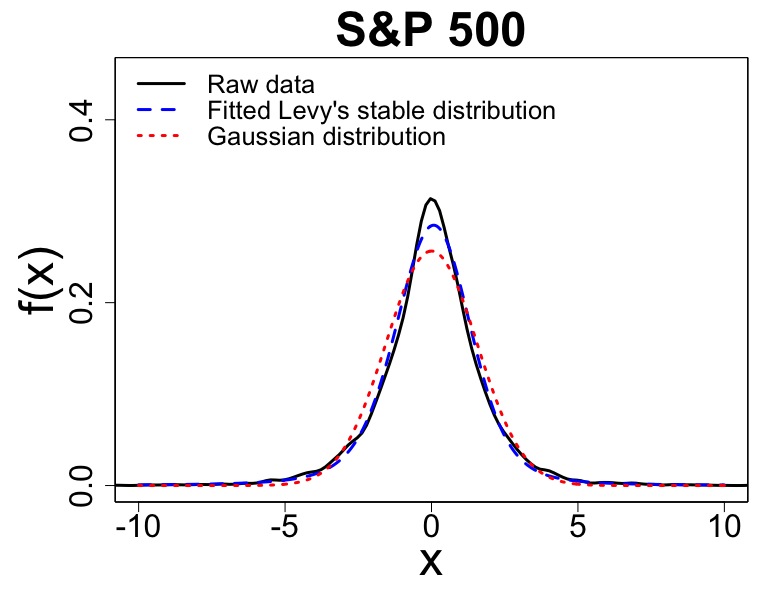}
\end{minipage}\\
\begin{minipage}{0.5\hsize}
\includegraphics[width=4.3cm,bb=0 0 767 598]{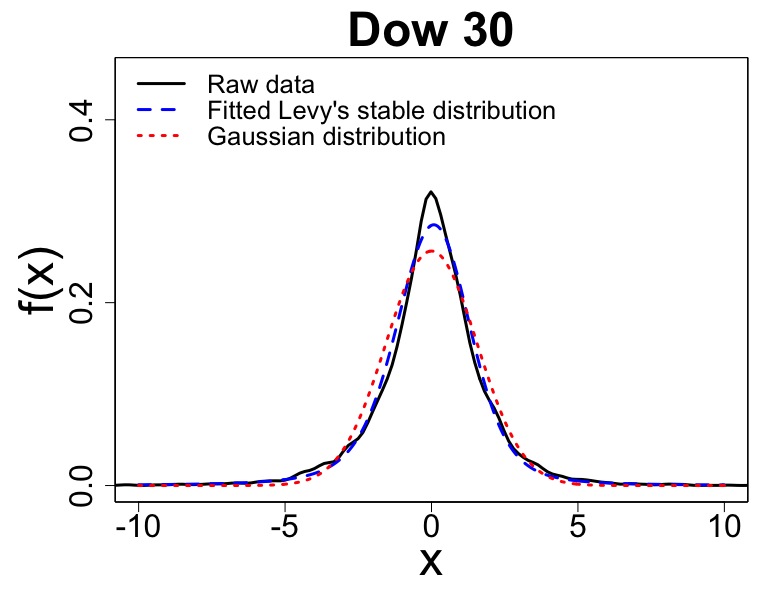}
\end{minipage}
\begin{minipage}{0.5\hsize}
\includegraphics[width=4.3cm,bb=0 0 767 598]{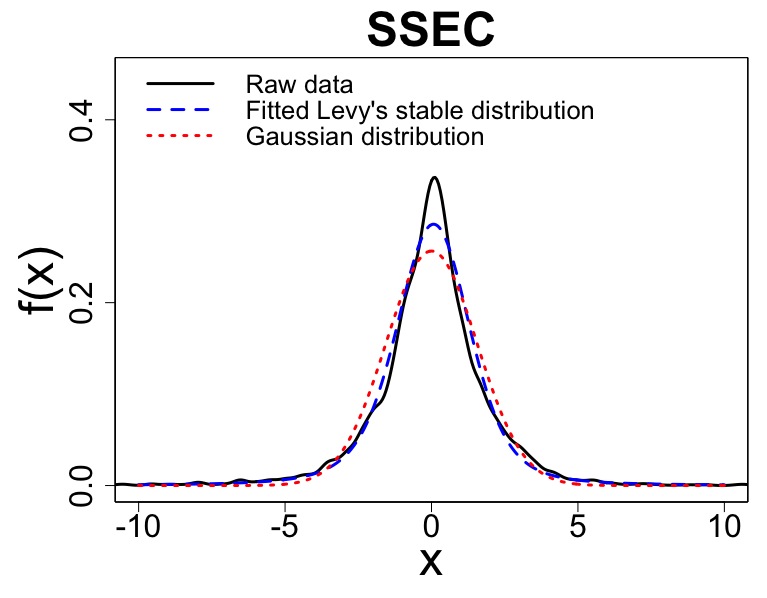}
\end{minipage}
\end{tabular}
\caption{Distribution of raw data, L\'evy's stable distribution with estimated parameters $(\alpha,\beta)$ and Gaussian distribution are compared.}
\label{fig:distribution}
\end{figure}

\begin{figure}[b]
\begin{tabular}{cc}
\begin{minipage}{0.5\hsize}
\includegraphics[width=4.3cm,bb=0 0 689 590]{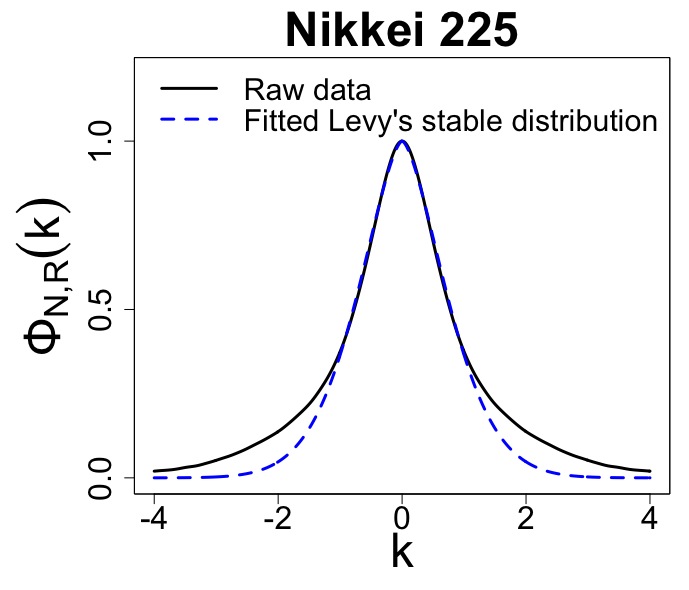}
\end{minipage}
\begin{minipage}{0.5\hsize}
\includegraphics[width=4.3cm,bb=0 0 689 590]{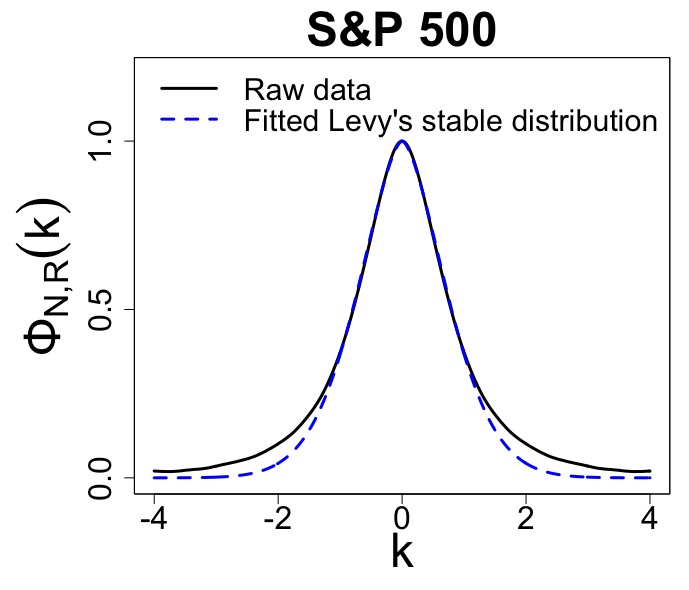}
\end{minipage}\\
\begin{minipage}{0.5\hsize}
\includegraphics[width=4.3cm,bb=0 0 689 590]{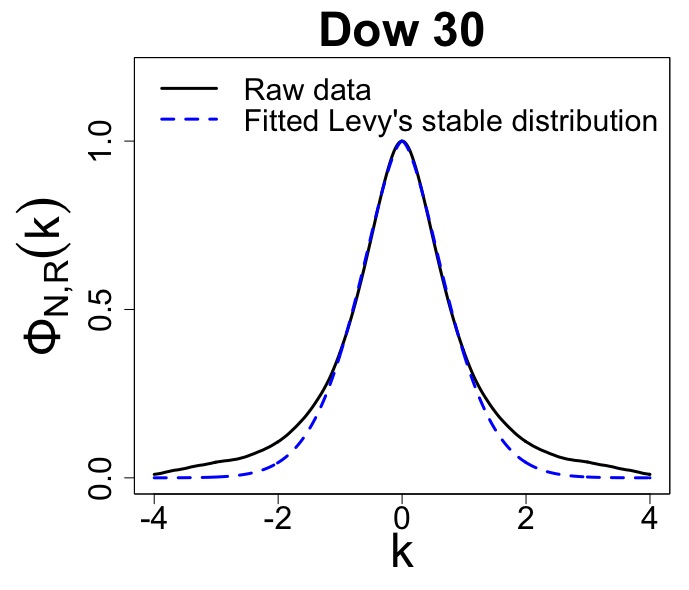}
\end{minipage}
\begin{minipage}{0.5\hsize}
\includegraphics[width=4.3cm,bb=0 0 689 590]{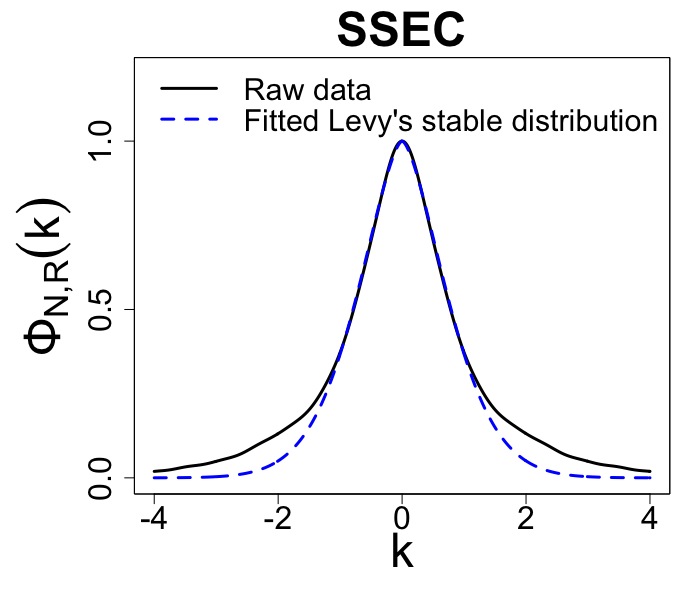}
\end{minipage}
\end{tabular}
\caption{Real part of characteristic function $\phi_R(k)$ of raw data and that of L\'evy's stable distribution with estimated parameters $(\alpha,\beta)$ are compared.}
\label{fig:cf_real}
\end{figure}

\begin{figure}[tb]
\begin{tabular}{cc}
\begin{minipage}{0.5\hsize}
\includegraphics[width=4.3cm,bb=0 0 689 590]{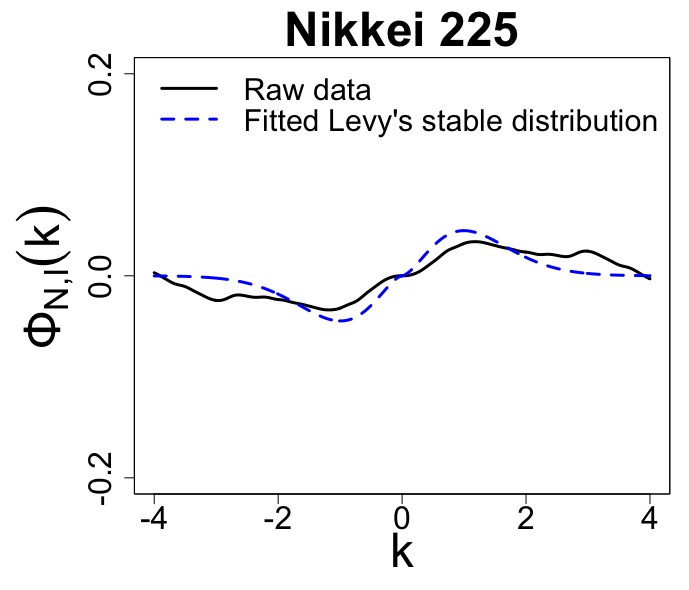}
\end{minipage}
\begin{minipage}{0.5\hsize}
\includegraphics[width=4.3cm,bb=0 0 689 590]{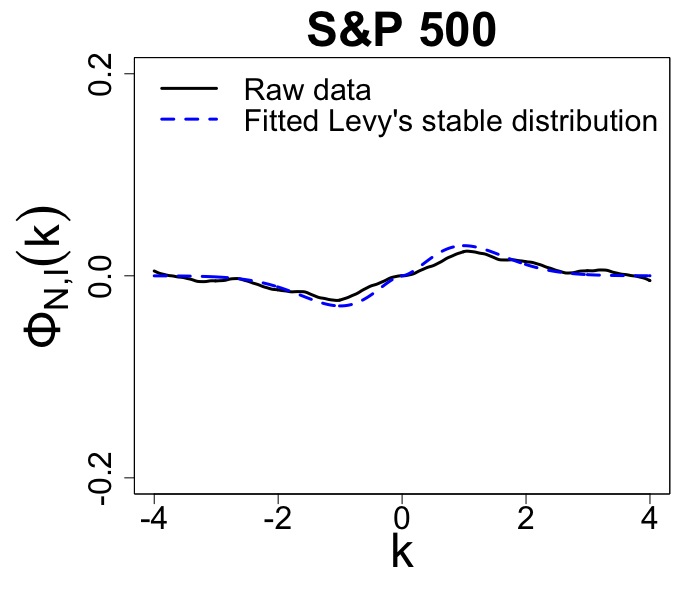}
\end{minipage}\\
\begin{minipage}{0.5\hsize}
\includegraphics[width=4.3cm,bb=0 0 689 590]{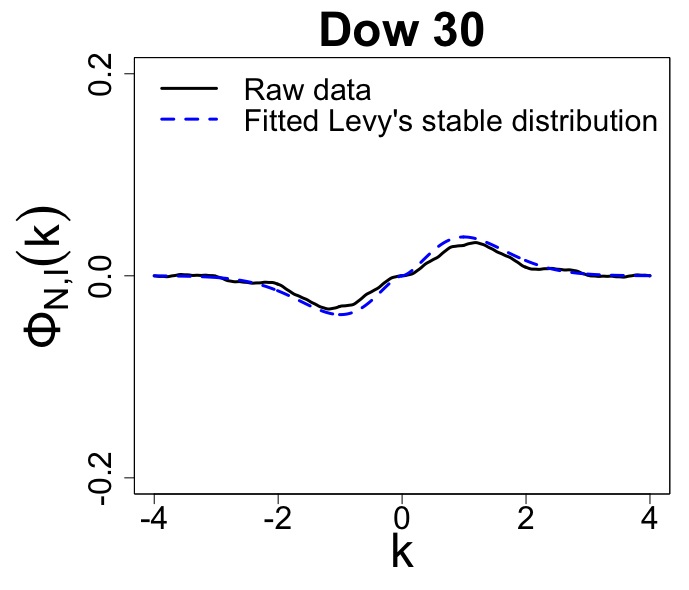}
\end{minipage}
\begin{minipage}{0.5\hsize}
\includegraphics[width=4.3cm,bb=0 0 689 590]{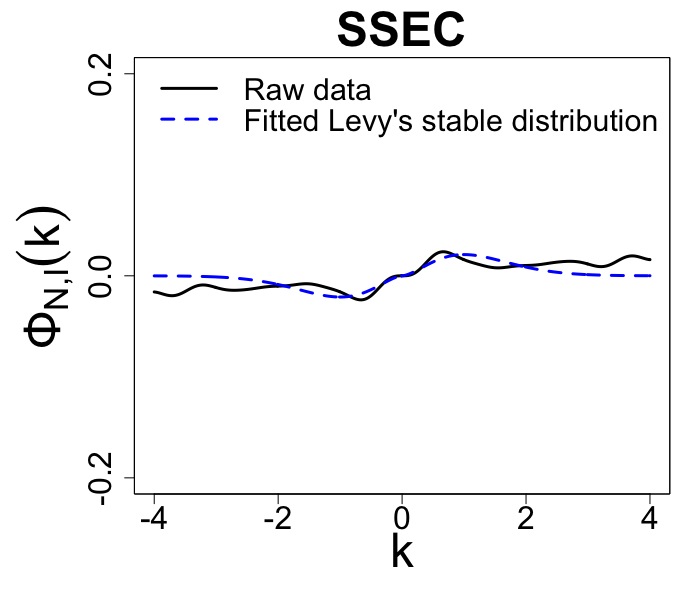}
\end{minipage}
\end{tabular}
\caption{Imaginary part of characteristic function $\phi_I(k)$ of raw data and that of L\'evy's stable distribution with estimated parameters $(\alpha,\beta)$ are compared.}
\label{fig:cf_imaginary}
\end{figure}

Fig. \ref{fig:distribution} compares the {\it probability} density functions of the standardized raw data, L\'evy's stable distribution with estimated parameters $(\alpha,\beta)$, and the Gaussian distribution. Figs. \ref{fig:cf_real} and \ref{fig:cf_imaginary} compare the {\it characteristic} functions. The parameters are estimated well, especially in the tail parts where the power-law influences. Note that $\beta$ always indicates the {\it negative} value, which means that the distribution has a non-zero skewness.

Next, let us see how the fitted parameters $(\alpha,\beta)$ change as a function of time. The parameters are estimated from $X(t-1000)$ to $X(t-1)$, which are the analysis results for the day $t$. That is to say, we move a window of $N=1000$ and compare the fluctuations of both the prices and the parameters $(\alpha,\beta)$.

\begin{figure}[t]
\includegraphics[width=8.6cm,bb=0 0 900 600]{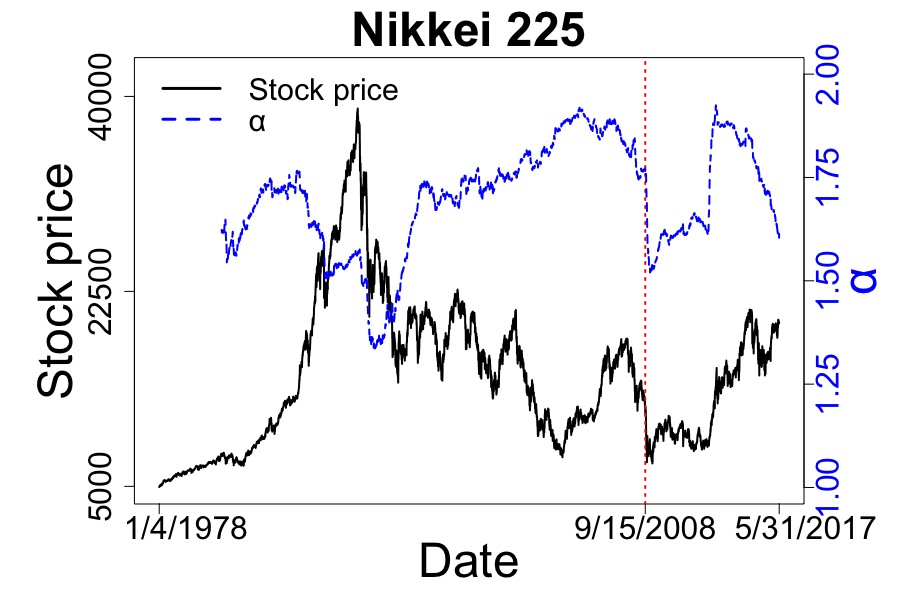}
\includegraphics[width=8.6cm,bb=0 0 900 600]{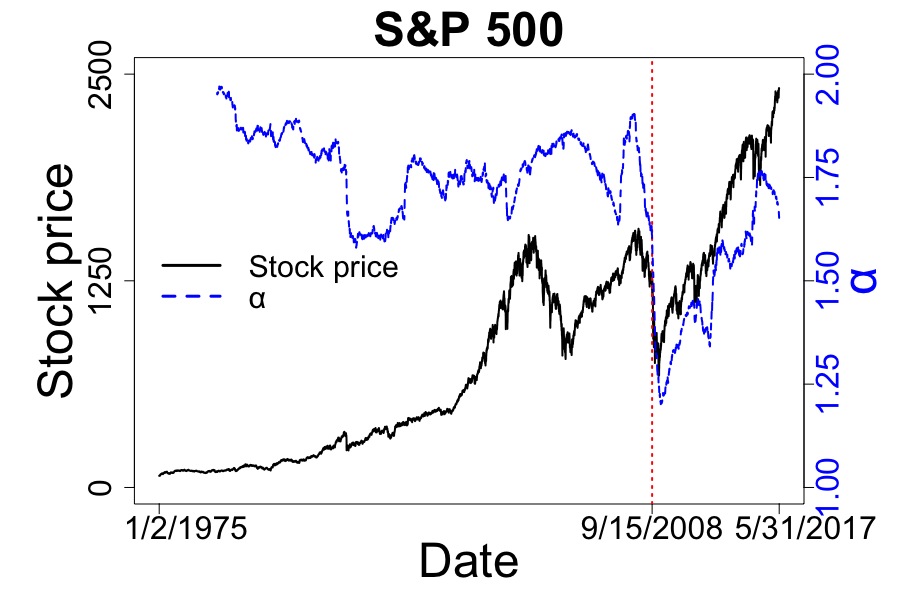}
\caption{Stock price and estimated parameter $\alpha$ are compared. Vertical dotted line corresponds to the date of the bankruptcy of Lehman Brothers.}
\label{fig:window_alpha}
\end{figure}

\begin{figure}[b]
\includegraphics[width=8.6cm,bb=0 0 900 600]{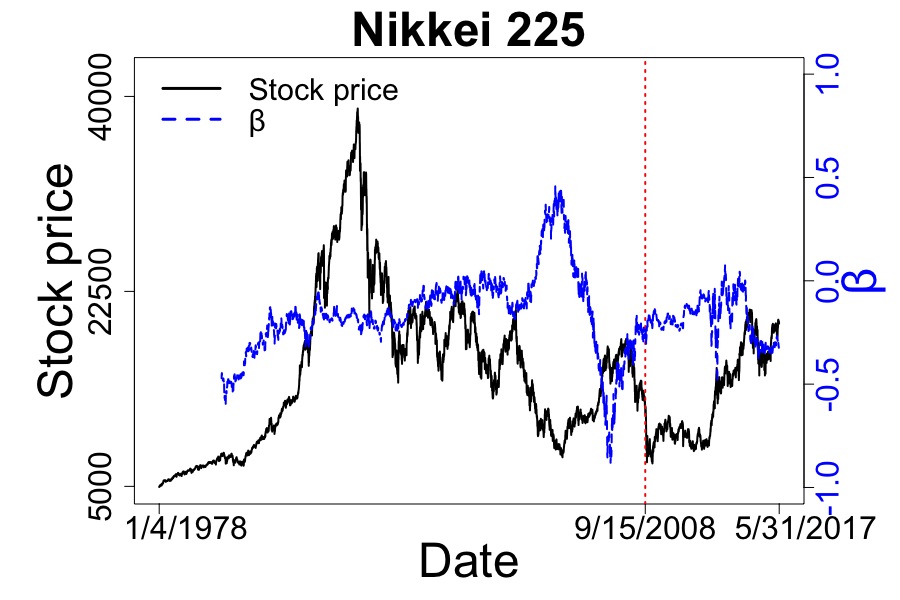}
\includegraphics[width=8.6cm,bb=0 0 900 600]{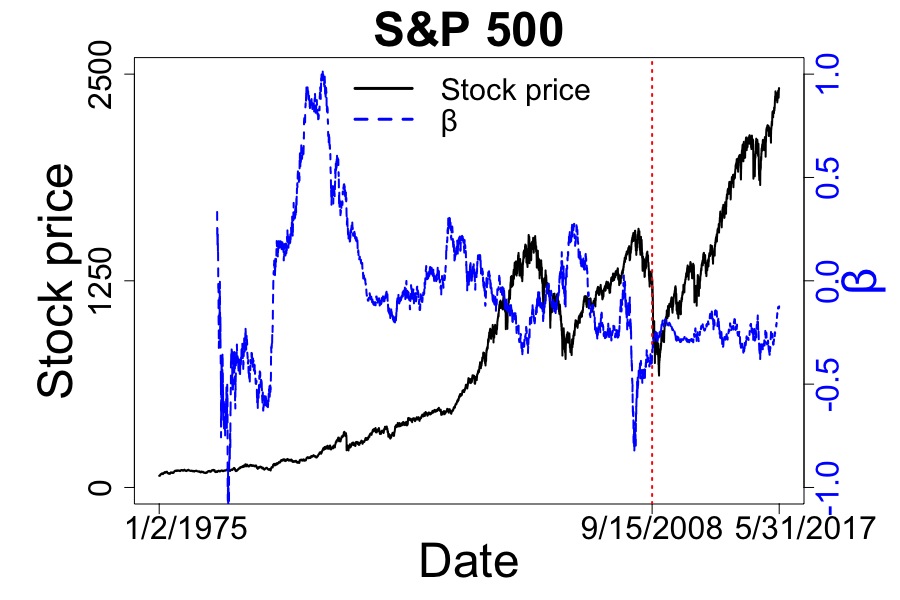}
\caption{Stock price and estimated parameter $\beta$ are compared. Vertical dotted line corresponds to the date of the bankruptcy of Lehman Brothers.}
\label{fig:window_beta}
\end{figure}

\begin{table*}
\caption{Local minimum values according to the financial crisis.}
\label{tab:minimum}
\begin{ruledtabular}
\begin{tabular}{ccccccc}
&\multicolumn{2}{c}{Price}&\multicolumn{2}{c}{$\alpha$}&\multicolumn{2}{c}{$\beta$}\\
Stock index&Value&Data number&Value&Data number&Value&Data number\\\hline
Nikkei 225&$7054.98$&$7905$&$1.516$&$7875$&$-0.886$&$7232$\\
S\&P 500&$676.53$&$8627$&$1.201$&$8656$&$-0.821$&$8199$\\
\end{tabular}
\end{ruledtabular}
\end{table*}

Figs. \ref{fig:window_alpha} and \ref{fig:window_beta} show the results of Nikkei 225 and S\&P 500. When looked at on a daily basis, the estimated parameters $(\alpha,\beta)$ fluctuate a lot around the average values shown in Table \ref{tab:stock}. Let us focus on the financial crisis which occurred in 2008, particularly when Lehman Brothers went bankrupt in September 15th, 2008. As the prices of both Nikkei 225 and S\&P 500 greatly crash, $\alpha$ shows a remarkable jump downwards, while $\beta$ is less affected.

Shown in  Fig. \ref{fig:window_alpha}, the index $\alpha$ shows a similar movement to the price. They start to go down from around 9/15/2008, then take the minimum values at almost the same time shown in Table \ref{tab:minimum}. Next, let us see the index $\beta$ in Fig. \ref{fig:window_beta}. The skewness parameter $\beta$ reflects the price fluctuation slightly shifted to the side, taking the minimum value before the price and $\alpha$. Accordingly, $\beta$ is more sensitive as if it had a short-time prediction of the price crashes such as the financial crisis.

\section{Conclusions}

In terms of the generalized central limit theorem, L\'evy's stable distribution is theoretically more suitable than the Gaussian distribution for fitting the log-returns of the stock markets. The stock prices with power-law tails would not converge to the Gaussian distribution, since the classical central limit theorem cannot be applied in this case.

The parameters $(\alpha,\beta)$ show a similar value regardless of the stock index. The stability parameter $\alpha$ of all the stock indices were around $\alpha=1.6$ which seems to be universal, and {\it lower} than the Gaussian distribution corresponding to $\alpha=2$.  As $\beta$ has a negative value, it is shown that the stock market has a {\it skewness}. Then, the parameters fluctuate by dividing the analyzing time-windows. There is a correlation between the price and $(\alpha,\beta)$, especially when the financial crisis occurred.

Stock prices often show larger fluctuations than expected before, and the probability distribution alters on a daily basis. In other words, stock markets are {\it instable}. This instability was not considered in the classical analysis, however, it should be incorporated in the models to prevent financial crisis from now on. As shown in this paper, the parameters of L\'evy's stable distribution could be the indicators of stock markets.

Although we used the data of daily stock prices in this paper, further development is expected by using the data with shorter intervals, such as the stock prices of every few minutes or even shorter with every few seconds.

\begin{acknowledgments}
The authors would like to thank Dr. Shin-itiro Goto, Kyoto University, for giving us insightful comments and carefully proofreading the paper.
\end{acknowledgments}

\nocite{*}
\bibliography{arxiv}

\begin{thebibliography}{27}%
\makeatletter
\providecommand \@ifxundefined [1]{%
 \@ifx{#1\undefined}
}%
\providecommand \@ifnum [1]{%
 \ifnum #1\expandafter \@firstoftwo
 \else \expandafter \@secondoftwo
 \fi
}%
\providecommand \@ifx [1]{%
 \ifx #1\expandafter \@firstoftwo
 \else \expandafter \@secondoftwo
 \fi
}%
\providecommand \natexlab [1]{#1}%
\providecommand \enquote  [1]{``#1''}%
\providecommand \bibnamefont  [1]{#1}%
\providecommand \bibfnamefont [1]{#1}%
\providecommand \citenamefont [1]{#1}%
\providecommand \href@noop [0]{\@secondoftwo}%
\providecommand \href [0]{\begingroup \@sanitize@url \@href}%
\providecommand \@href[1]{\@@startlink{#1}\@@href}%
\providecommand \@@href[1]{\endgroup#1\@@endlink}%
\providecommand \@sanitize@url [0]{\catcode `\\12\catcode `\$12\catcode
  `\&12\catcode `\#12\catcode `\^12\catcode `\_12\catcode `\%12\relax}%
\providecommand \@@startlink[1]{}%
\providecommand \@@endlink[0]{}%
\providecommand \url  [0]{\begingroup\@sanitize@url \@url }%
\providecommand \@url [1]{\endgroup\@href {#1}{\urlprefix }}%
\providecommand \urlprefix  [0]{URL }%
\providecommand \Eprint [0]{\href }%
\providecommand \doibase [0]{http://dx.doi.org/}%
\providecommand \selectlanguage [0]{\@gobble}%
\providecommand \bibinfo  [0]{\@secondoftwo}%
\providecommand \bibfield  [0]{\@secondoftwo}%
\providecommand \translation [1]{[#1]}%
\providecommand \BibitemOpen [0]{}%
\providecommand \bibitemStop [0]{}%
\providecommand \bibitemNoStop [0]{.\EOS\space}%
\providecommand \EOS [0]{\spacefactor3000\relax}%
\providecommand \BibitemShut  [1]{\csname bibitem#1\endcsname}%
\let\auto@bib@innerbib\@empty
\bibitem [{\citenamefont {Black}\ and\ \citenamefont {Scholes}(1973)}]{bs1973}%
  \BibitemOpen
  \bibfield  {author} {\bibinfo {author} {\bibfnamefont {F.}~\bibnamefont
  {Black}}\ and\ \bibinfo {author} {\bibfnamefont {M.}~\bibnamefont
  {Scholes}},\ }\href@noop {} {\bibfield  {journal} {\bibinfo  {journal} {J.
  Polit. Econ.}\ }\textbf {\bibinfo {volume} {81}},\ \bibinfo {pages} {637}
  (\bibinfo {year} {1973})}\BibitemShut {NoStop}%
\bibitem [{\citenamefont {Markowitz}(1952)}]{markowitz1952}%
  \BibitemOpen
  \bibfield  {author} {\bibinfo {author} {\bibfnamefont {H.}~\bibnamefont
  {Markowitz}},\ }\href@noop {} {\bibfield  {journal} {\bibinfo  {journal} {J.
  Financ.}\ }\textbf {\bibinfo {volume} {7}},\ \bibinfo {pages} {77} (\bibinfo
  {year} {1952})}\BibitemShut {NoStop}%
\bibitem [{\citenamefont {Merton}(1971)}]{merton1971}%
  \BibitemOpen
  \bibfield  {author} {\bibinfo {author} {\bibfnamefont {R.}~\bibnamefont
  {Merton}},\ }\href@noop {} {\bibfield  {journal} {\bibinfo  {journal} {J.
  Econ. Theory}\ }\textbf {\bibinfo {volume} {3}},\ \bibinfo {pages} {373}
  (\bibinfo {year} {1971})}\BibitemShut {NoStop}%
\bibitem [{\citenamefont {Mandelbrot}(1963)}]{mandelbrot1963}%
  \BibitemOpen
  \bibfield  {author} {\bibinfo {author} {\bibfnamefont {B.}~\bibnamefont
  {Mandelbrot}},\ }\href@noop {} {\bibfield  {journal} {\bibinfo  {journal} {J.
  Bus.}\ }\textbf {\bibinfo {volume} {36}},\ \bibinfo {pages} {394} (\bibinfo
  {year} {1963})}\BibitemShut {NoStop}%
\bibitem [{\citenamefont {Fama}(1965)}]{fama1965}%
  \BibitemOpen
  \bibfield  {author} {\bibinfo {author} {\bibfnamefont {E.~F.}\ \bibnamefont
  {Fama}},\ }\href@noop {} {\bibfield  {journal} {\bibinfo  {journal} {J.
  Bus.}\ }\textbf {\bibinfo {volume} {38}},\ \bibinfo {pages} {34} (\bibinfo
  {year} {1965})}\BibitemShut {NoStop}%
\bibitem [{\citenamefont {D.~A.~Hsu}\ and\ \citenamefont
  {Wichern}(1974)}]{miller1974}%
  \BibitemOpen
  \bibfield  {author} {\bibinfo {author} {\bibfnamefont {R.~B.~M.}\
  \bibnamefont {D.~A.~Hsu}}\ and\ \bibinfo {author} {\bibfnamefont {D.~W.}\
  \bibnamefont {Wichern}},\ }\href@noop {} {\bibfield  {journal} {\bibinfo
  {journal} {J. Am. Stat. Assoc.}\ }\textbf {\bibinfo {volume} {69}},\ \bibinfo
  {pages} {108} (\bibinfo {year} {1974})}\BibitemShut {NoStop}%
\bibitem [{\citenamefont {Mantegna}\ and\ \citenamefont
  {Stanley}(1995)}]{mantegna1995}%
  \BibitemOpen
  \bibfield  {author} {\bibinfo {author} {\bibfnamefont {R.~N.}\ \bibnamefont
  {Mantegna}}\ and\ \bibinfo {author} {\bibfnamefont {H.~E.}\ \bibnamefont
  {Stanley}},\ }\href@noop {} {\bibfield  {journal} {\bibinfo  {journal}
  {Nature}\ }\textbf {\bibinfo {volume} {376}},\ \bibinfo {pages} {46}
  (\bibinfo {year} {1995})}\BibitemShut {NoStop}%
\bibitem [{\citenamefont {Liu}\ \emph {et~al.}(1999)\citenamefont {Liu},
  \citenamefont {Gopikrishnan}, \citenamefont {Cizeau}, \citenamefont {Meyer},
  \citenamefont {Peng},\ and\ \citenamefont {Stanley}}]{liu1999}%
  \BibitemOpen
  \bibfield  {author} {\bibinfo {author} {\bibfnamefont {Y.}~\bibnamefont
  {Liu}}, \bibinfo {author} {\bibfnamefont {P.}~\bibnamefont {Gopikrishnan}},
  \bibinfo {author} {\bibfnamefont {P.}~\bibnamefont {Cizeau}}, \bibinfo
  {author} {\bibfnamefont {M.}~\bibnamefont {Meyer}}, \bibinfo {author}
  {\bibfnamefont {C.~K.}\ \bibnamefont {Peng}}, \ and\ \bibinfo {author}
  {\bibfnamefont {H.~E.}\ \bibnamefont {Stanley}},\ }\href@noop {} {\bibfield
  {journal} {\bibinfo  {journal} {Phys.\ Rev. E}\ }\textbf {\bibinfo {volume}
  {60}},\ \bibinfo {pages} {1390} (\bibinfo {year} {1999})}\BibitemShut
  {NoStop}%
\bibitem [{\citenamefont {Gopikrishnan}\ \emph {et~al.}(1999)\citenamefont
  {Gopikrishnan}, \citenamefont {Plerou}, \citenamefont {Amaral}, \citenamefont
  {Meyer},\ and\ \citenamefont {Stanley}}]{gopikrishnan1999}%
  \BibitemOpen
  \bibfield  {author} {\bibinfo {author} {\bibfnamefont {P.}~\bibnamefont
  {Gopikrishnan}}, \bibinfo {author} {\bibfnamefont {V.}~\bibnamefont
  {Plerou}}, \bibinfo {author} {\bibfnamefont {L.~A.~N.}\ \bibnamefont
  {Amaral}}, \bibinfo {author} {\bibfnamefont {M.}~\bibnamefont {Meyer}}, \
  and\ \bibinfo {author} {\bibfnamefont {H.~E.}\ \bibnamefont {Stanley}},\
  }\href@noop {} {\bibfield  {journal} {\bibinfo  {journal} {Phys.\ Rev. E}\
  }\textbf {\bibinfo {volume} {60}},\ \bibinfo {pages} {5305} (\bibinfo {year}
  {1999})}\BibitemShut {NoStop}%
\bibitem [{\citenamefont {Plerou}\ \emph {et~al.}(1999)\citenamefont {Plerou},
  \citenamefont {Gopikrishnan}, \citenamefont {Amaral}, \citenamefont {Meyer},\
  and\ \citenamefont {Stanley}}]{plerou1999}%
  \BibitemOpen
  \bibfield  {author} {\bibinfo {author} {\bibfnamefont {V.}~\bibnamefont
  {Plerou}}, \bibinfo {author} {\bibfnamefont {P.}~\bibnamefont
  {Gopikrishnan}}, \bibinfo {author} {\bibfnamefont {L.~A.~N.}\ \bibnamefont
  {Amaral}}, \bibinfo {author} {\bibfnamefont {M.}~\bibnamefont {Meyer}}, \
  and\ \bibinfo {author} {\bibfnamefont {H.~E.}\ \bibnamefont {Stanley}},\
  }\href@noop {} {\bibfield  {journal} {\bibinfo  {journal} {Phys.\ Rev. E}\
  }\textbf {\bibinfo {volume} {60}},\ \bibinfo {pages} {6519} (\bibinfo {year}
  {1999})}\BibitemShut {NoStop}%
\bibitem [{\citenamefont {Plerou}\ \emph {et~al.}(2000)\citenamefont {Plerou},
  \citenamefont {Gopikrishnan}, \citenamefont {Amaral}, \citenamefont
  {Gabaix},\ and\ \citenamefont {Stanley}}]{plerou2000}%
  \BibitemOpen
  \bibfield  {author} {\bibinfo {author} {\bibfnamefont {V.}~\bibnamefont
  {Plerou}}, \bibinfo {author} {\bibfnamefont {P.}~\bibnamefont
  {Gopikrishnan}}, \bibinfo {author} {\bibfnamefont {L.~A.~N.}\ \bibnamefont
  {Amaral}}, \bibinfo {author} {\bibfnamefont {X.}~\bibnamefont {Gabaix}}, \
  and\ \bibinfo {author} {\bibfnamefont {H.~E.}\ \bibnamefont {Stanley}},\
  }\href@noop {} {\bibfield  {journal} {\bibinfo  {journal} {Phys.\ Rev. E}\
  }\textbf {\bibinfo {volume} {62}},\ \bibinfo {pages} {3023} (\bibinfo {year}
  {2000})}\BibitemShut {NoStop}%
\bibitem [{\citenamefont {Gopikrishnan}\ \emph {et~al.}(2000)\citenamefont
  {Gopikrishnan}, \citenamefont {Plerou}, \citenamefont {Gabaix},\ and\
  \citenamefont {Stanley}}]{gopikrishnan2000}%
  \BibitemOpen
  \bibfield  {author} {\bibinfo {author} {\bibfnamefont {P.}~\bibnamefont
  {Gopikrishnan}}, \bibinfo {author} {\bibfnamefont {V.}~\bibnamefont
  {Plerou}}, \bibinfo {author} {\bibfnamefont {X.}~\bibnamefont {Gabaix}}, \
  and\ \bibinfo {author} {\bibfnamefont {H.~E.}\ \bibnamefont {Stanley}},\
  }\href@noop {} {\bibfield  {journal} {\bibinfo  {journal} {Phys.\ Rev. E}\
  }\textbf {\bibinfo {volume} {62}},\ \bibinfo {pages} {4493} (\bibinfo {year}
  {2000})}\BibitemShut {NoStop}%
\bibitem [{\citenamefont {Rachev}\ and\ \citenamefont
  {Mittnik}(2000)}]{rachev2000}%
  \BibitemOpen
  \bibfield  {author} {\bibinfo {author} {\bibfnamefont {S.~T.}\ \bibnamefont
  {Rachev}}\ and\ \bibinfo {author} {\bibfnamefont {S.}~\bibnamefont
  {Mittnik}},\ }\href@noop {} {\emph {\bibinfo {title} {Stable Paretian Models
  in Finance}}}\ (\bibinfo  {publisher} {Wiley},\ \bibinfo {year}
  {2000})\BibitemShut {NoStop}%
\bibitem [{\citenamefont {Gabaix}\ \emph {et~al.}(2003)\citenamefont {Gabaix},
  \citenamefont {Gopikrishnan}, \citenamefont {Plerou},\ and\ \citenamefont
  {Stanley}}]{stanley2003}%
  \BibitemOpen
  \bibfield  {author} {\bibinfo {author} {\bibfnamefont {X.}~\bibnamefont
  {Gabaix}}, \bibinfo {author} {\bibfnamefont {P.}~\bibnamefont
  {Gopikrishnan}}, \bibinfo {author} {\bibfnamefont {V.}~\bibnamefont
  {Plerou}}, \ and\ \bibinfo {author} {\bibfnamefont {H.~E.}\ \bibnamefont
  {Stanley}},\ }\href@noop {} {\bibfield  {journal} {\bibinfo  {journal}
  {Nature}\ }\textbf {\bibinfo {volume} {423}},\ \bibinfo {pages} {267}
  (\bibinfo {year} {2003})}\BibitemShut {NoStop}%
\bibitem [{\citenamefont {Weron}(2001)}]{weron2001}%
  \BibitemOpen
  \bibfield  {author} {\bibinfo {author} {\bibfnamefont {R.}~\bibnamefont
  {Weron}},\ }\href@noop {} {\bibfield  {journal} {\bibinfo  {journal} {Int. J.
  Mod. Phys. C}\ }\textbf {\bibinfo {volume} {12}},\ \bibinfo {pages} {209}
  (\bibinfo {year} {2001})}\BibitemShut {NoStop}%
\bibitem [{\citenamefont {Fama}\ and\ \citenamefont {Roll}(1971)}]{fama1971}%
  \BibitemOpen
  \bibfield  {author} {\bibinfo {author} {\bibfnamefont {E.~F.}\ \bibnamefont
  {Fama}}\ and\ \bibinfo {author} {\bibfnamefont {R.}~\bibnamefont {Roll}},\
  }\href@noop {} {\bibfield  {journal} {\bibinfo  {journal} {J. Am. Stat.
  Assoc.}\ }\textbf {\bibinfo {volume} {66}},\ \bibinfo {pages} {331} (\bibinfo
  {year} {1971})}\BibitemShut {NoStop}%
\bibitem [{\citenamefont {McCulloch}(1986)}]{mcculloch1986}%
  \BibitemOpen
  \bibfield  {author} {\bibinfo {author} {\bibfnamefont {J.~H.}\ \bibnamefont
  {McCulloch}},\ }\href@noop {} {\bibfield  {journal} {\bibinfo  {journal}
  {Commun. Stat. Simulat.}\ }\textbf {\bibinfo {volume} {15}},\ \bibinfo
  {pages} {1109} (\bibinfo {year} {1986})}\BibitemShut {NoStop}%
\bibitem [{\citenamefont {Shao}\ and\ \citenamefont
  {Nikias}(1993)}]{nikias1993}%
  \BibitemOpen
  \bibfield  {author} {\bibinfo {author} {\bibfnamefont {M.}~\bibnamefont
  {Shao}}\ and\ \bibinfo {author} {\bibfnamefont {C.~L.}\ \bibnamefont
  {Nikias}},\ }\href@noop {} {\bibfield  {journal} {\bibinfo  {journal} {P.
  IEEE}\ }\textbf {\bibinfo {volume} {81}},\ \bibinfo {pages} {986} (\bibinfo
  {year} {1993})}\BibitemShut {NoStop}%
\bibitem [{\citenamefont {Ma}\ and\ \citenamefont {Nikias}(1995)}]{nikias1995}%
  \BibitemOpen
  \bibfield  {author} {\bibinfo {author} {\bibfnamefont {X.~Y.}\ \bibnamefont
  {Ma}}\ and\ \bibinfo {author} {\bibfnamefont {C.~L.}\ \bibnamefont
  {Nikias}},\ }\href@noop {} {\bibfield  {journal} {\bibinfo  {journal} {IEEE
  T. Signal Proces.}\ }\textbf {\bibinfo {volume} {43}},\ \bibinfo {pages}
  {2884} (\bibinfo {year} {1995})}\BibitemShut {NoStop}%
\bibitem [{\citenamefont {Kuruo\v{g}lu}(2001)}]{kuruoglu2001}%
  \BibitemOpen
  \bibfield  {author} {\bibinfo {author} {\bibfnamefont {E.~E.}\ \bibnamefont
  {Kuruo\v{g}lu}},\ }\href@noop {} {\bibfield  {journal} {\bibinfo  {journal}
  {IEEE T. Signal Proces.}\ }\textbf {\bibinfo {volume} {49}},\ \bibinfo
  {pages} {2192} (\bibinfo {year} {2001})}\BibitemShut {NoStop}%
\bibitem [{\citenamefont {Koutrouvelis}(1980)}]{koutrouvelis1980}%
  \BibitemOpen
  \bibfield  {author} {\bibinfo {author} {\bibfnamefont {I.~A.}\ \bibnamefont
  {Koutrouvelis}},\ }\href@noop {} {\bibfield  {journal} {\bibinfo  {journal}
  {J. Am. Stat. Assoc.}\ }\textbf {\bibinfo {volume} {75}},\ \bibinfo {pages}
  {918} (\bibinfo {year} {1980})}\BibitemShut {NoStop}%
\bibitem [{\citenamefont {Gnedenko}\ and\ \citenamefont
  {Kolmogorov}(1954)}]{gclt1954}%
  \BibitemOpen
  \bibfield  {author} {\bibinfo {author} {\bibfnamefont {B.~V.}\ \bibnamefont
  {Gnedenko}}\ and\ \bibinfo {author} {\bibfnamefont {A.~N.}\ \bibnamefont
  {Kolmogorov}},\ }\href@noop {} {\emph {\bibinfo {title} {Limit Distributions
  for Sums of Independent Random Variables}}}\ (\bibinfo  {publisher}
  {Addison-Wesley},\ \bibinfo {year} {1954})\BibitemShut {NoStop}%
\bibitem [{\citenamefont {Shintani}\ and\ \citenamefont
  {Umeno}()}]{shintani2017}%
  \BibitemOpen
  \bibfield  {author} {\bibinfo {author} {\bibfnamefont {M.}~\bibnamefont
  {Shintani}}\ and\ \bibinfo {author} {\bibfnamefont {K.}~\bibnamefont
  {Umeno}},\ }\href@noop {} {\bibinfo  {journal} {arXiv:1702.02826}\
  }\BibitemShut {NoStop}%
\bibitem [{\citenamefont {Halmos}(1956)}]{halmos1956}%
  \BibitemOpen
\bibfield  {journal} {  }\bibfield  {author} {\bibinfo {author} {\bibfnamefont
  {P.~R.}\ \bibnamefont {Halmos}},\ }\href@noop {} {\emph {\bibinfo {title}
  {Lectures on Ergodic Theory}}}\ (\bibinfo  {publisher} {Amer Mathematical
  Society},\ \bibinfo {year} {1956})\BibitemShut {NoStop}%
\bibitem [{\citenamefont {Nemytskii}\ and\ \citenamefont
  {Stepanov}(1960)}]{nemytskii1960}%
  \BibitemOpen
  \bibfield  {author} {\bibinfo {author} {\bibfnamefont {V.~V.}\ \bibnamefont
  {Nemytskii}}\ and\ \bibinfo {author} {\bibfnamefont {V.~V.}\ \bibnamefont
  {Stepanov}},\ }\href@noop {} {\emph {\bibinfo {title} {Qualitative Theory of
  Differential Equations}}}\ (\bibinfo  {publisher} {Princeton University
  Press},\ \bibinfo {year} {1960})\BibitemShut {NoStop}%
\bibitem [{\citenamefont {Umeno}(2016)}]{umeno2016}%
  \BibitemOpen
  \bibfield  {author} {\bibinfo {author} {\bibfnamefont {K.}~\bibnamefont
  {Umeno}},\ }\href@noop {} {\bibfield  {journal} {\bibinfo  {journal} {NOLTA}\
  }\textbf {\bibinfo {volume} {7}},\ \bibinfo {pages} {14} (\bibinfo {year}
  {2016})}\BibitemShut {NoStop}%
\bibitem [{\citenamefont {Voit}(2005)}]{voit2005}%
  \BibitemOpen
  \bibfield  {author} {\bibinfo {author} {\bibfnamefont {J.}~\bibnamefont
  {Voit}},\ }\href@noop {} {\emph {\bibinfo {title} {The Statistical Mechanics
  of Financial Markets}}}\ (\bibinfo  {publisher} {Springer},\ \bibinfo {year}
  {2005})\BibitemShut {NoStop}%
\end{thebibliography}%

\end{document}